**RESEARCH**

# Read mapping on de Bruijn graphs

Antoine Limasset[1*], Bastien Cazaux[2,3], Eric Rivals[2,3] and Pierre Peterlongo[1]

*Correspondence:
antoine.limasset@irisa.fr
[1]IRISA Inria Rennes Bretagne Atlantique, GenScale team, Campus de Beaulieu, 35042 Rennes, France
Full list of author information is available at the end of the article

**Abstract**

**Background** Next Generation Sequencing (NGS) has dramatically enhanced our ability to sequence genomes, but not to assemble them. In practice, many published genome sequences remain in the state of a large set of contigs. Each contig describes the sequence found along some path of the assembly graph, however, the set of contigs does not record all the sequence information contained in that graph. Although many subsequent analyses can be performed with the set of contigs, one may ask whether mapping reads on the contigs is as informative as mapping them on the paths of the assembly graph. Currently, one lacks practical tools to perform mapping on such graphs.

**Results** Here, we propose a formal definition of mapping on a de Bruijn graph, analyse the problem complexity which turns out to be NP-complete, and provide a practical solution. We propose a pipeline called *GGMAP* (Greedy Graph MAPping). Its novelty is a procedure to map reads on branching paths of the graph, for which we designed a heuristic algorithm called *BGREAT* (de Bruijn Graph REAd mapping Tool). For the sake of efficiency, *BGREAT* rewrites a read sequence as a succession of unitigs sequences. *GGMAP* can map millions of reads per CPU hour on a de Bruijn graph built from a large set of human genomic reads. Surprisingly, results show that up to 22% more reads can be mapped on the graph but not on the contig set.

**Conclusions** Although mapping reads on a de Bruijn graph is complex task, our proposal offers a practical solution combining efficiency with an improved mapping capacity compared to assembly-based mapping even for complex eukaryotic data.

## 1 Background

Next Generation Sequencing technologies (NGS) have drastically accelerated the generation of sequenced genomes. However, these technologies remain unable to provide a single sequence per chromosome. Instead, they produce a large and redundant set of reads, with each read being a piece of the whole genome. Because of this redundancy, it is possible to detect overlaps between reads and to assemble them together in order to reconstruct the target genome sequence.

Even today, assembling reads remains a complex task for which no single piece of software performs consistently well [1]. The assembly problem itself has been shown to be computationally difficult, more precisely NP-hard [2]. Practical limitations arise both from the structure of genomes (repeats longer than reads cannot be correctly resolved) and from the sequencing biases (non-uniform coverage and sequencing errors). Applied solutions represent the sequence of the reads in an assembly graph: the labels along a path of the graph encode a sequence. Currently, most assemblers rely on two types of graphs: either the de Bruijn graph (DBG) for the short reads produced by the second generation of sequencing technologies [3], or for long reads the overlap graph (which was introduced in the Celera Assembler [4]) and variants thereof, like the string graph [5]. Then, the assembly algorithm explores the graph using heuristics, selects some paths and outputs their sequences. Due to these heuristics, the set of sequences obtained, called contigs, is biased and fragmented because of complex patterns in the graph that are generated by sequencing errors, and genomic variants and repeats. The set of contigs is rarely satisfactory and is usually post-processed, for instance, by discarding short contigs.

The most frequent computational task for analyzing a set of reads is mapping them on a reference genome. Numerous tools are available to map reads when the reference genome has the form of a set of sequences (e.g. BWA [6] and Bowtie [7]). The goal of mapping on a finished genome sequence is to say whether a sequence can be aligned to this genome, and in this case, at which location(s). This is mostly done with a heuristic (semi-global) alignment procedure that authorizes a small edit or Hamming distance between the read and genome sequences. Read mapping process suffers from regions of low mappability [8]. Repeated genomic regions may not be mapped precisely since the reads mapping on these regions have multiple matches. When a genome is represented as a graph, the mappability issue is reduced, as occurrences of each repeated region are factorized, limiting the problem of multiple matches of reads.

When the reference is not a finished genome sequence, but a redundant set of contigs, the situation differs. The mapping may correctly determine whether the read is found in the genome, but multiple locations may for instance not be sufficient to conclude whether several true locations exist. Conversely, an unfruitful mapping of a read may be due to an incomplete assembly or to the removal of some contigs during post-processing. In such cases, we argue it may be interesting to consider the assembly graph as a (less biased and/or more complete) reference instead of the set of contigs. Then mapping on the paths of this graph is needed to complement mapping on set of contigs. This motivates the design and implementation of BGREAT.

In this context, we explore the problem of mapping reads on a graph. Aligning or mapping sequences on sequence graphs (a generic term meaning a graph representing sequences along its paths) has already been explored in the literature in different application contexts: assembly, read correction, or metagenomics.

In the context of assembly, once a DBG has been built, mapping the reads back to the graph can help in eliminating unsupported paths or in computing the coverage of edges. To our knowledge, no practical solution has been designed for this task. Cerulean assembler [9] mentions this possibility, but only uses regular alignment on assembled sequences. Allpaths-LG [10] also performs a similar task to resolve repeats using long noisy reads from third generation sequencing techniques. Its procedure is not generic enough to suit the mapping of any read set on a DBG. From the theoretical view point, the question is related to the NP-hard *read-threading* problem (also termed *Eulerian superpath problem* [11, 2]), which consists in finding a read coherent path in the DBG (a path that can be represented as a sequence of reads as defined in [5]). The assembler called SPADES [12] threads the reads against the DBG by keeping track of the paths used during construction, which requires a substantial amount of memory. Here, we propose a more general problem, termed *De Bruijn Graph Read Mapping Problem* (DBGRMP), as we aim at mapping to a graph any source of NGS reads, either those reads used for building the graph or other reads.

Recently, the hybrid error correction of long reads using short reads has become a critical step to leverage the third generation of sequencing technologies. The error corrector

LoRDEC [13] builds the DBG of the short reads, and then aligns each long read against the paths of the DBG by computing their edit distance using a dynamic programming algorithm (which is slow for our purposes). For shorts reads correction, several tools that evaluate the *k*-mer spectrum of reads to correct the sequencing errors use a probabilistic or an exact representation of a DBG as a reference [14, 15].

In the context of metagenomics, Wang et al [16] have estimated the taxonomic composition of a metagenomics sample by mapping reads on a DBG representing several genomes of closely-related bacterial species. In fact, the graph collapses similar regions of these genomes and avoids redundant mapping. Their tool maps the read using BWA on the sequence resulting from the random concatenation of unitigs of the DBG. Hence, a read cannot align over several successive nodes of the graph (ER: il y a un pb ce n'est pas vrai). Similarly, several authors have proposed to store related genomes into a single, less repetitive, DBG [17, 18, 19]. However, most of these tools are efficient only when applied to very closely related sequences that result in flat graphs. The *BlastGraph* tool [19], is specifically dedicated to the mapping of reads on graphs, but is unusable on real world graphs (see Results section).

Here, we formalize the mapping of reads on a De Bruijn graph and show that it is NP-complete. Then we present the pipeline *GGMAP* and dwell on *BGREAT*, a new tool which enables to map reads on branching paths of the DBG (Section 2.2). For the sake of efficiency, *BGREAT* adopts a heuristic algorithm that scales up to huge sequencing data sets. In Section 3, we evaluate *GGMAP* in terms of mapping capacity and of efficiency, and compare it to mapping on assembled contigs. Finally, we discuss the limitations and advantages the of *GGMAP* and give some directions of future work (Section 4).

## 2 Methods

We formally define the problem of mapping reads on a DBG and investigate its complexity (Section 2.1). Besides, we propose a pipeline called *GGMAP* to map short reads on a representation of a DBG (Section 2.2). This pipeline includes *BGREAT*, a new algorithm mapping sequences on branching paths of the graph (Section 2.3).

### 2.1 Complexity of mapping reads on the paths of a DBG

In this section, we present the formal problem we aim to solve and prove its intractability. First, we introduce preliminary definitions, then formalize the problem of mapping reads on paths of a DBG, called the De Bruijn Graph Read Mapping Problem (DBGRMP), and finally prove it is NP-complete. Our starting point is the well-known Hamiltonian Path Problem (HPP); we apply several reductions to prove the hardness of DBGRMP.

**Definition 1** (de Bruijn graph)

*Given a set of strings $S = \{r_1, r_2, ..., r_n\}$ on an alphabet $\Sigma$ and an integer $k \geq 2$, the de Bruijn graph of order k of S ($dBG_k(S)$) is a directed graph $(V,A)$ where:*
$V = \{d \in \Sigma^k \mid \exists\, i \in \{1,\ldots,n\}$ *such that d is a substring of $r_i \in S\}$, and*
$A = \{(d,d') \mid$ *if the suffix of length $k-1$ of d is a prefix of $d'\}$.*

**Definition 2** (Walk and Path of a directed graph)

*Let G be a directed graph.*
- *A* walk *of G is an alternating sequence of nodes and connecting edges of G.*

- A path *of G is a walk of G without repeated node.*
- A Hamiltonian path *is a path that that visits each node of G exactly once.*

**Definition 3** (Sequence generated by a walk in a *dBG$_k$*)  *Let G be a de Bruijn graph of order k. A walk of G composed of l nodes $(v_1,\ldots,v_l)$ generates a sequence of length $k + l - 1$ obtained by the concatenation of $v_1$ with the last character of $v_2$, of $v_3$ ,..., of $v_l$.*

We define the *de Bruijn Graph Read Mapping Problem* (*DBGRMP*) as follows:

**Definition 4** (De Bruijn Graph Read Mapping Problem)  *Given*
- *S, a set of strings over $\Sigma$,*
- *k, an integer such that $k \geq 2$,*
- *$q := q_1 \ldots q_{|q|}$ a word of $\Sigma^*$ such that $|q| \geq k$,*
- *a cost function $F : \Sigma \times \Sigma \to \mathbb{N}$, and*
- *a threshold $t \in \mathbb{N}$,*

*decide whether there exists a path of the $dBG_k(S)$ composed of $|q| - k + 1$ nodes (generating a word $m := m_1 \ldots m_{|q|} \in \Sigma^{|q|}$) such that the cost $C(m,q) := \sum_{i=1}^{|q|} F(m_i, q_i) \leq t$.*

We recall the definition of the *Hamiltonian Path Problem* (*HPP*), which is NP-complete [20].

**Definition 5** (*Hamiltonian Path Problem* (*HPP*))  *Given a directed graph G, the HPP consists in deciding whether there exists a Hamiltonian path of G.*

To prove the NP-completeness of DBGRMP we introduce two intermediate problems. The first problem is a variant of the Asymmetrical Travelling Salesman Problem.

**Definition 6** (*Fixed Length Asymmetric Travelling Salesman Problem* (*FLATSP*))  *Let*
- *l be an integer,*
- *$G := (V,A,c)$ be a directed graph whose edges are labeled with a non-negative integer cost (given by the function $c : A \to \mathbb{N}$),*
- *$t \in \mathbb{N}$ be a threshold.*

*FLATSP consists in deciding whether there exists a path $p := (v_1,\ldots,v_l)$ of G composed of l nodes whose cost $c(p) := \sum_{j=1}^{l-1} c((v_j, v_{j+1}))$ satisfies $c(p) \leq t$.*

We consider the restriction of FLATSP to instances having a unit cost function (i.e., where $c(a) = 1$ for any $a \in A$) and where $l$ equals both the threshold and the number of nodes in $V$. This restriction makes FLATSP very similar to HPP, and the hardness result quite natural.

**Proposition 1**  *FLATSP is NP-complete even when restricted to instances with a unit cost function and satisfying $l = |V| = t$.*

*Proof*  We reduce HPP to an instance of FLATSP where the cost function $c$ simply counts the edges in the path, and where the path length $l$ equals the threshold $t$ and the number of nodes in $V$.

Let $G = (V, A)$ be a directed graph, which is an instance of HPP. Let $H = (V, A, c : A \to \{1\})$, and $l := |V|$ and $t := l$. Thus $(H, l, t)$ is an instance of FLATSP.

Let us now show that there is an equivalence between the existence of a Hamiltonian path in $G$ and the existence of a path $p = (v_1, \ldots, v_l)$ of $H$ such that $c(p) \leq t$. Assume that $G$ has a Hamiltonian path $p$. In this case, $p$ is also a path in $H$ of length $|V|$, and then the cost of $p$ equals its length, i.e. $c(p) = \sum_{i=1}^{|V|} 1 = |V|$. Hence, there exists a path $p$ of $H$ such that $c(p) \leq t = |V|$.

Assume that there exists a path $p = (v_1, \ldots, v_{|V|})$ of $H$ such that $c(p) \leq t$. As $p$ is a path it has no repeated nodes, and as by assumption $l = |V|$, one gets that $p$ is a Hamiltonian path of $H$, and thus also a Hamiltonian path of $G$, since $G$ and $H$ share the same set of nodes and edges. □

The second intermediate problem is called the *Read Graph Mapping Problem* (*GRMP*) and is defined below. It formalizes the mapping on a general sequence graph. Hence, DB-GRMP is a specialization of GRMP, since it considers the case of the de Bruijn graph.

**Definition 7** (Graph Read Mapping Problem)  *Given*
- *a directed graph $G = (V, A, x)$, whose edges are labeled by symbols of the alphabet ($x : A \to \Sigma$),*
- $q := q_1 \ldots q_{|q|}$ *a word of $\Sigma^*$,*
- *a cost function $F : \Sigma \times \Sigma \to \mathbb{N}$,*
- *a threshold $t \in \mathbb{N}$,*

*GRMP consists in deciding whether there exists a path $p := (v_1, \ldots, v_{|q|+1})$ of $G$ composed of $|q| + 1$ nodes, which generates a word $m := m_1 \ldots m_{|q|} \in \Sigma^{|q|}$ such that $m_i := x((v_i, v_{i+1}))$, and which satisfies $\sum_{i=1}^{|q|} F(m_i, q_i) \leq t$. Here, $m$ is called the word generated by $p$.*

**Proposition 2**  *GRMP is NP-complete.*

*Proof* We reduce FLATSP to GRMP.

Let $(G = (V, A, c : A \to \mathbb{N}), l \in \mathbb{N}, t \in \mathbb{N})$ be an instance of FLATSP. Let $\Sigma = \{y_1, \ldots, y_{|\Sigma|}\}$ an alphabet larger than the largest value of $c(A)$, and let $s$ be the application such that $s : \{0, \ldots, |\Sigma|\} \to \Sigma$ and such that for each $i$ in $\{0, \ldots, |\Sigma|\}$, $s(i) = y_i$. Let $H = (V, A, x := s \circ c)$ and let $\alpha$ be a letter that does not belong to $\Sigma$, let $q = \alpha^{l-1}$ and $F$ such that for each $i$ in $\{0, \ldots, |\Sigma|\}$, $F(\alpha, y_i) = i$. Thus, we obtain $|q| = l - 1$.

Now, let us show that there is an equivalence between the existence of a path $p = (v_1, \ldots, v_l)$ of $G$ such that $c(p) \leq t$ and the existence of a path $p' = (u_1, \ldots, u_{|q|+1})$ of $H$ composed of $|q| + 1$ nodes, which generates a word $m = m_1 \ldots m_{|q|}$ of $\Sigma^{|q|}$, where each $m_j = x((u_j, u_{j+1}))$, and such that $\sum_{j=1}^{|q|} F(m_j, q_j) \leq t$. Assume that there exists a path $p = (v_1, \ldots, v_l)$ of $G$ such that $c(p) \leq t$. By definition, $p$ is a path in $H$. Let $m$ be the word generated by $p$. Thus we have $\sum_{j=1}^{|q|} F(m_j, q_j) = \sum_{j=1}^{l-1} F(m_j, \alpha) = \sum_{j=1}^{l-1} c((v_j, v_{j+1})) \leq t$.

Now, suppose that there exists a path $p' = (u_1, \ldots, u_{|q|+1})$ of $H$ composed of $|q| + 1$ nodes, which generates a word $m = m_1 \ldots m_{|q|}$ of $\Sigma^{|q|}$, where each $m_j = x((u_j, u_{j+1}))$, and such that $\sum_{j=1}^{|q|} F(m_j, q_j) \leq t$. By the construction of $H$, $p'$ is a path in $G$ of length $|q| + 1 = l$. Hence, we obtain $\sum_{j=1}^{l-1} c((u_j, u_{j+1})) = \sum_{j=1}^{|q|} F(m_j, \alpha) = \sum_{j=1}^{l-1} F(m_j, q_j) \leq t$. □

**Theorem 1**  *DBGRMP is NP-complete.*

Figure 1 illustrates the gadget used in the proof of Theorem 1. Basically, the gadget creates a DBG node (a word) formed by concatenating the labels of the two preceding edges in the original graph.

*Proof* Let us now reduce GRMP to DBGRMP.

Let $(G := (V, A, x : A \to \Sigma), q \in \Sigma^*, F : \Sigma \times \Sigma \to \mathbb{N}, t \in \mathbb{N})$ be an instance of GRMP. Let \$ and $\Delta$ be two distinct letters that do not belong to $\Sigma$, and let $\Sigma' := \Sigma \cup \{\$, \Delta\}$. Let $V'$ be a set of words of length 2 defined by

$$
\begin{aligned}
V' := &\{\alpha_i \beta_j \mid x(i,j) = \alpha \text{ and } \exists\, l \in V \text{ such that } x(j,l) = \beta\} & \textbf{set 1} \\
&\bigcup \{\Delta_i \$_i \mid \exists\, j \in V, \text{ such that } x(i,j) = \alpha \text{ and } \nexists\, l \in V \text{ such that } (l,i) \in A\} & \textbf{set 2} \\
&\bigcup \{\$_i \alpha_i \mid \exists\, j \in V, \text{ such that } x(i,j) = \alpha \text{ and } \nexists\, l \in V \text{ such that } (l,i) \in A\}. & \textbf{set 3}
\end{aligned}
\tag{1}
$$

Any letter of a word in $V'$ is a symbol of $\Sigma'$ numbered by a node of $V$. Moreover, if that symbol is taken from $V$ then it labels an edge of $A$ that goes out a node, say $i$, of $V$, and the number associated to that symbol is $i$. In fact, $V'$ is the union of three sets (see Equation 1): set 1 considers the cases of an edge of $A$ labeled $\alpha$ followed by an edge labeled $\beta$, sets 2 and 3 contain the cases of an edge of $A$ labeled $\alpha$ that is not preceded by another edge of $A$; for each such edge one creates two words: $\Delta_i \$_i$ in set 2 and $\$_i \alpha_i$ in set 3.

Let $H$ be the 2-dBG of $V'$; note that $\Sigma'$ is the alphabet of the words of $V'$. Now let $z$ be the application from $V'$ to $\Sigma$ that for any $\alpha_i$ of $V'$ satisfies $z(\alpha_i) = \alpha$. (Note that in this equation, the right term is a shortcut meaning the symbol of $\alpha_i$ without its numbering $i$; this shortcut is used only for the sake of legibility, but can be properly written with a heavier notation). Let $F' : \Sigma' \times \Sigma \to \mathbb{N}$ be the application such that $\forall (\alpha_i, \beta) \in \Sigma' \times \Sigma$, $F'(\alpha_i, \beta) = F(z(\alpha_i), \beta) = F(\alpha, \beta)$.

Let us show that this reduction is a bijection that transforms a positive instance of GRMP into a positive instance of DBGRMP. Assume there exists a path $p := (v_1, \ldots, v_{|q|+1})$ of $G$ which generates a word $m = m_1 \ldots m_{|q|} \in \Sigma^{|q|}$ satisfying $m_i = x((v_i, v_{i+1}))$ and such that $\sum_{i=1}^{|q|} F(m_i, q_i) \leq t$. We show that there exists a path $p'$ of $G'$ which generates a word $m' = m'_1 \ldots m'_{|q|} \in \Sigma'^{|q|}$ such that $\sum_{i=1}^{|q|} F'(m'_i, q_i) \leq t$.

We build the path $p'$ as the "concatenation" of two paths, denoted $p'_{start}$ and $p'_{end}$, that we define below. Let $\gamma_j := x((v_j, v_{j+1}))_{v_j} = (m_j)_{v_j}$ for all $j$ between 1 and $|q|$. One has that $\gamma_j \in \Sigma'$. Now, let

$$
p'_{start} := \begin{cases}
\left(x((v_{l'}, v_l))_{v_{l'}} x((v_l, v_1))_{v_l},\ x((v_l, v_1))_{v_l} x((v_1, v_2))_{v_1}\right) \\
\quad \text{if } \exists\, l, l' \in V \text{ such that } (l, 1) \in A \text{ and } (l', l) \in A \\
\left(\$_{v_l} x((v_l, v_1))_{v_l},\ x((v_l, v_1))_{v_l} x((v_1, v_2))_{v_1}\right) \\
\quad \text{if } \exists\, l \in V \text{ such that } (l, 1) \in A \text{ and } \nexists\, l' \in V \text{ such that } (l', l) \in A \\
\left(\Delta_{v_1} \$_{v_1},\ \$_{v_1} x((v_1, v_2))_{v_1}\right) \\
\quad \text{otherwise.}
\end{cases}
$$

and let

$$
p'_{end} := \left(\gamma_1 \gamma_2,\ \ldots,\ \gamma_{|q|-1} \gamma_{|q|}\right).
$$

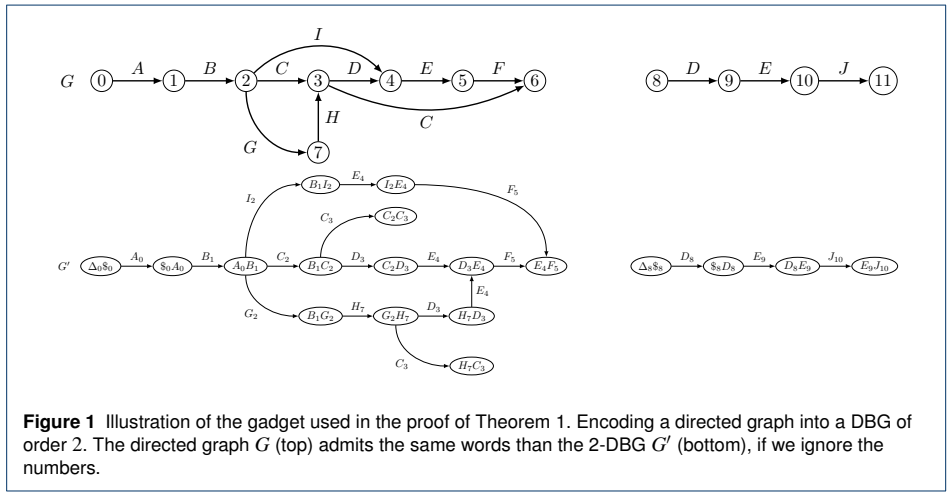

**Figure 1** Illustration of the gadget used in the proof of Theorem 1. Encoding a directed graph into a DBG of order 2. The directed graph $G$ (top) admits the same words than the 2-DBG $G'$ (bottom), if we ignore the numbers.

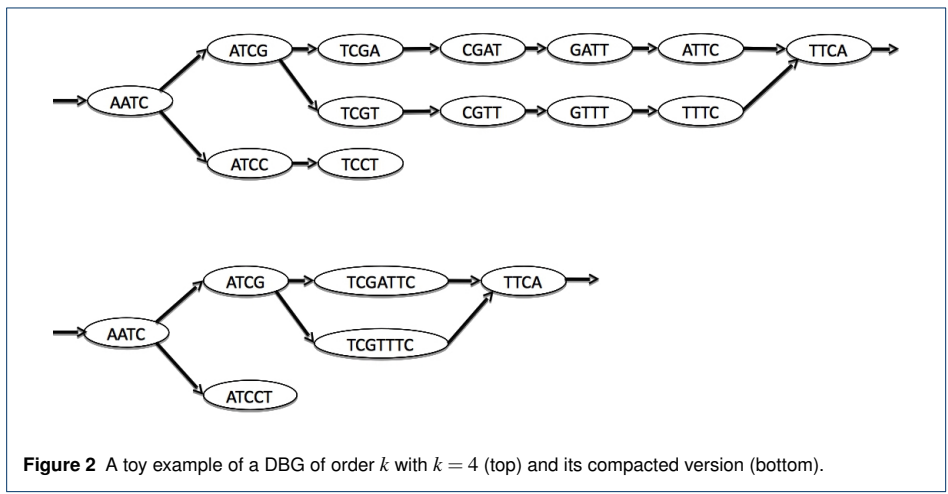

**Figure 2** A toy example of a DBG of order $k$ with $k = 4$ (top) and its compacted version (bottom).

Let $m'$ denote the word generated by $p'$. Clearly, one sees that $m' = (m_1)_{v_1} \ldots (m_{|q|})_{v_{|q|}}$, and since $m_i = z((m'_i)_{v_i})$, one gets that $z(m') = m$ and $\sum_{i=1}^{|q|} F'(m'_i, q_i) = \sum_{i=1}^{|q|} F(m_i, q_i) \leq t$.

In the other direction, the proof is similar since our construction is a bijection. □

### 2.2 *GGMAP*: a method to map reads on de Bruijn Graph

We propose a practical solution for solving DBGRMP. We consider the case of short (hundred of base pairs) reads with a low error rate (1% of substitution), which is a good approximation of widely used NGS reads. Since errors are mostly substitutions, mapping is computed using the Hamming distance. Our solution is designed for mapping on a compacted de Bruijn graph (CDBG) any set of short reads, either those used to build the graph or reads from another individual or species. We recall that a CDBG is representation of a DBG in which each non branching path is merged into a single node. The sequence of each node is called a *unitig*. Figure 2 shows a DBG and the associated CDBG.

In a CDBG, the nodes are not necessarily $k$-mers, words of length $k$, but *unitigs*, with some unitigs being longer than reads. Thus, while mapping on a CDBG, one distinguishes between two mapping situations: **i/** the reads mapping completely on a unitig of the graph, and **ii/** the reads whose mapping spans two or more unitigs. For the latter, we say that the read *maps on a branching path of the graph*.

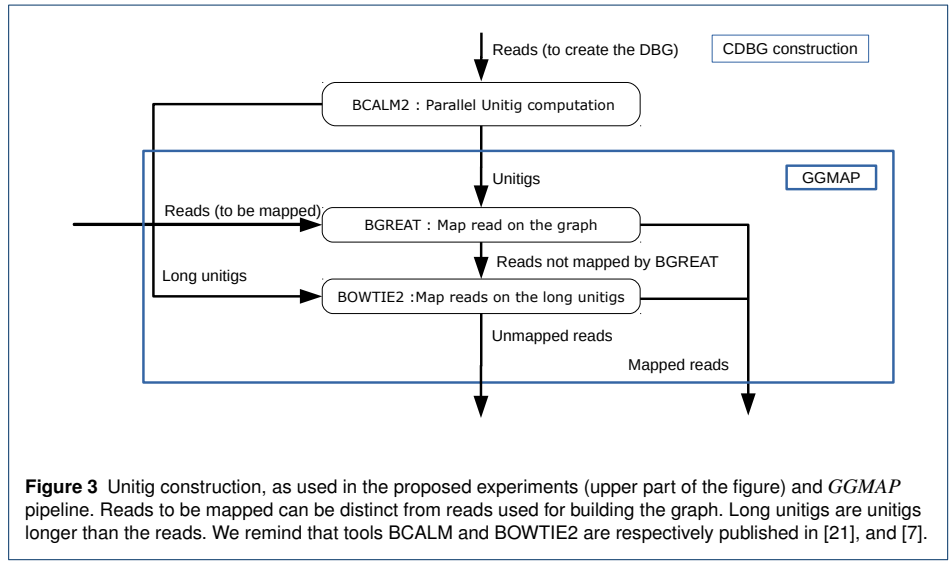

**Figure 3** Unitig construction, as used in the proposed experiments (upper part of the figure) and *GGMAP* pipeline. Reads to be mapped can be distinct from reads used for building the graph. Long unitigs are unitigs longer than the reads. We remind that tools BCALM and BOWTIE2 are respectively published in [21], and [7].

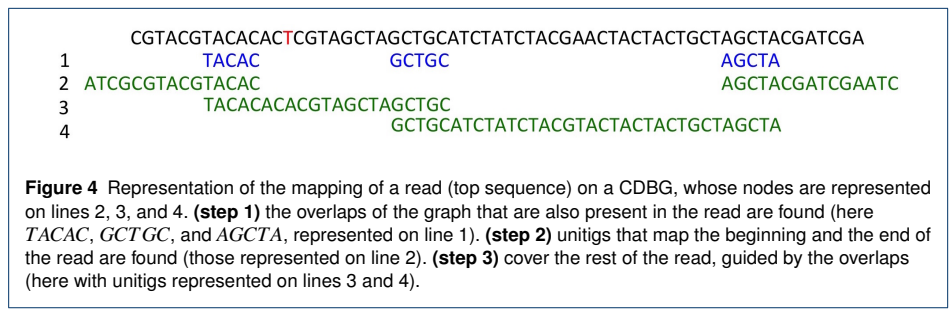

**Figure 4** Representation of the mapping of a read (top sequence) on a CDBG, whose nodes are represented on lines 2, 3, and 4. **(step 1)** the overlaps of the graph that are also present in the read are found (here *TACAC*, *GCTGC*, and *AGCTA*, represented on line 1). **(step 2)** unitigs that map the beginning and the end of the read are found (those represented on line 2). **(step 3)** cover the rest of the read, guided by the overlaps (here with unitigs represented on lines 3 and 4).

Taking advantage of the extensive research carried out for mapping reads on flat strings, *GGMAP* uses Bowtie2 [7] to map the reads on the unitigs. In addition, *GGMAP* integrates our proposed new tool, called *BGREAT*, for mapping reads on branching paths of the CDBG. Figure. 3 provides an overview of the pipeline.

*GGMAP* takes as inputs a query set of reads and a reference DBG. To avoid including sequencing errors in the DBG, we construct the reference DBG after filtering out all $k$-mers whose coverage lies below a user-defined threshold $c$. This error removal step is a classical preprocessing step that is performed in $k$-mer based assemblers. The unitigs of the CDBG are computed using *BCALM* 2 (the parallel version of *BCALM* [21]), using the $k$-mers having a coverage $\geq c$. *GGMAP* uses such a set of unitigs as DBG.

We now propose a detailed description of *BGREAT*.

### 2.3 *BGREAT*: mapping reads on branching paths of the CDBG

As previously mentioned, *BGREAT* is designed for mapping reads on branching paths of a CDBG, using reasonable resources both in terms of time and memory. Our approach follows the usual "seed and extend" paradigm. More generally, the proposed implementation applies heuristic schemes, both regarding the indexing and the alignment phases.

*2.3.1 Indexing heuristic*

We remind that our algorithm maps reads that span at least two distinct unitigs. Such mapped reads inevitably traverse one or more DBG edge(s). In a CDBG, edges are represented by the prefix and suffix of size $k-1$ of each unitig. We call such sequences the *overlaps*. In order to limit the index size and the computation time, our algorithm indexes only overlaps that are later used as seeds. Those overlaps are good anchors for several reasons: they are long enough ($k-1$) to be selective, they cannot be shared by more than eight unitigs (four starting and four ending with the overlap), and a CDBG usually has a reasonable number of unitigs and then of overlaps. For instance, the CDBG in our experiment with human data has 70 million unitigs and 87 million overlaps for 3 billion $k$-mers). In our implementation, the index is a minimal perfect hash table indicating for each overlap the unitig(s) starting or ending with this $(k-1)$-mer. Using a minimal perfect hash function limits the memory footprint, while keeping efficient query times (see Table 3).

---

**Data**: Read *r*, Integer *n*
**for** *the n first overlaps of r* **do**
    Find a path *begin* that map the begin of *r*
    **if** *begin* found **then**
        **for** *the n last overlaps of r* **do**
            Find a path *end* that maps the end of *r*
            **if** *end* found **then**
                Find (in a greedy way) a path *cover* that map the read from *begin* to *end*
                **if** *cover* found **then**
                      write path;
                      **return**

**Algorithm 1:** Greedy algorithm for mapping a read on multiple unitigs once the potential overlaps present in the read have been detected.

*2.3.2 Read alignment*

Given a read, each of its $k-1$-mers is used to query the index. The index detects which $k-1$-mers represent an overlap of the CDBG. An example of a read together with the matched unitigs are displayed on Figure. 4. Once the overlaps and their corresponding unitigs have been computed, the alignment of the read is performed from left to right as presented in Algorithm 1. Given an overlap position *i* on the read, the unitigs starting with this overlap are aligned to the sequence of the read starting from position *i*. The best alignment is recorded. In addition, to improve speed, if one of the at most four unitigs ending with the same overlap is the next overlap detected on the read, then this unitig is tested first, and if the alignment contains less mismatch than the user defined threshold, the other unitigs are not considered. Note that this optimization does not apply for the first and last overlaps of a read.

This mapping procedure is performed only if the two extremities of the read are mapped by two unitigs. The extreme overlaps of the read enables BGREAT to quickly filter out unmappable reads. For doing this, the first (resp. last) overlap of the read is used to align the read to the first (resp. last) unitig. Note that, as polymorphism exists between the read and the graph, some of the overlaps present on the read may be spurious. In this case the alignment fails, and the algorithm continues with the next (resp. previous) overlap. At most *n* alignment failures are authorized in each direction. If a read cannot be anchored neither on the left, nor on the right, it is considered as not aligned to the graph.

| CDBG Id | Reads Id | $k$ | $c$ | Number of unitigs | Mean length of unitigs |
|---|---|---|---|---|---|
| E.coli | SRR959239 | 31 | 3 | 42,843 | 134 |
| C.elegans_norm | SRR065390 | 31 | 3 | 1,627,335 | 93 |
| C.elegans_cpx | SRR065390 | 21 | 2 | 8,273,338 | 34 |
| Human | SRR345593 SRR345594 | 31 | 10 | 69,932,343 | 70 |

**Table 1** CDBG used in this study. *C.elegans*_cpx and *C.elegans*_norm are two distinct graphs, constructed using the same read set from *C.elegans* genome. The suffixes *norm* and *cpx* respectively stand for "normal" (using $c = 3$ and $k = 31$) and for "complex" (using a low threshold $c = 2$ and small value $k = 21$).

Note that the whole approach is greedy: given two or more possible choices, the best one is chosen and backtracking is excluded. This results in a linear time mapping process, since each position in the read can lead to a maximum of four comparisons, and the algorithm continues as long as the cumulated number of mismatches remains below the user defined threshold. Because of heuristics, a read may be unmapped or wrongly mapped for any of the following reasons.

- All overlaps on which the read should map contain errors, in this case the read is not anchored or only badly anchored and thus not mapped.
- The *n* first or *n* last overlaps of the read are spurious, in this case the *begin* or *end* is not found and the read is not mapped. By default and in all experiments $n = 2$.
- The greedy choices made during the path selection are wrong.

We implemented *BGREAT* as a dependence-free tool in C++ available at github.com/Malfoy/BGREAT.

## 3 Results

Beforehand we give details about the data sets (Subsection 3.1), then we perform several evaluations of *GGMAP* and of *BGREAT*. First, we compare graph mapping to mapping on the contigs resulting from an assembly (Subsection 3.2). Second, we assess how many reads are mapped on branching paths vs on unitigs (Subsection 3.3). Third, we evaluate the efficiency of BGREAT in both terms of throughput and scalability (Subsection 3.4), then assess the quality of the mapping itself (Subsection 3.5). All *BGREAT* alignments were performed authorizing up to two mismatches.

There are very few published tools to compare *GGMAP* with. Indeed, we found only one published tool, called *BlastGraph* [19], which was designed for mapping reads on a DBG. However, on our simplest data set coming from the *E.coli* genome (see Table 1), *BlastGraph* crashed after ≈ 124h of computation. Thus, *BlastGraph* was not further investigated here.

### 3.1 Data sets and CDBG construction

For our experiments we used publicly available Illumina read data sets from species of increasing complexity: from the bacterium *E.coli*, the worm *C.elegans*, and from Human. Detailed information about the data sets are given in Table 1 of the Additional File (identifiers, read length, read numbers, and coverages – from 70x to 112x–).

For each of these three data sets, we generated a CDBG using BCALM. From the *C.elegans* read set, we additionally generated an artificially complex graph, by using small *k* and *c* values (respectively 21 and 2). This particular graph, called *C.elegans_cpx*, contains lot of small unitigs. We used it to assess situations of highly complex and/or low quality sequencing data. The characteristics of the CDBG obtained on each of these data sets are given in Table 1.

| Set | % mapped on contigs | % mapped on CDBG |
|---|---|---|
| *E.coli* | 95.57 | 97.16 |
| *C.elegans_norm* | 80,60 | 93,24 |
| *C.elegans_cpx* | 56,33 | 89,15 |
| Human | 63,16 | 85,70 |

**Table 2** Percentage of mapped reads, either mapping on contigs (here obtained thank to the Minia assembler) or mapping on CDBG with *GGMAP*.

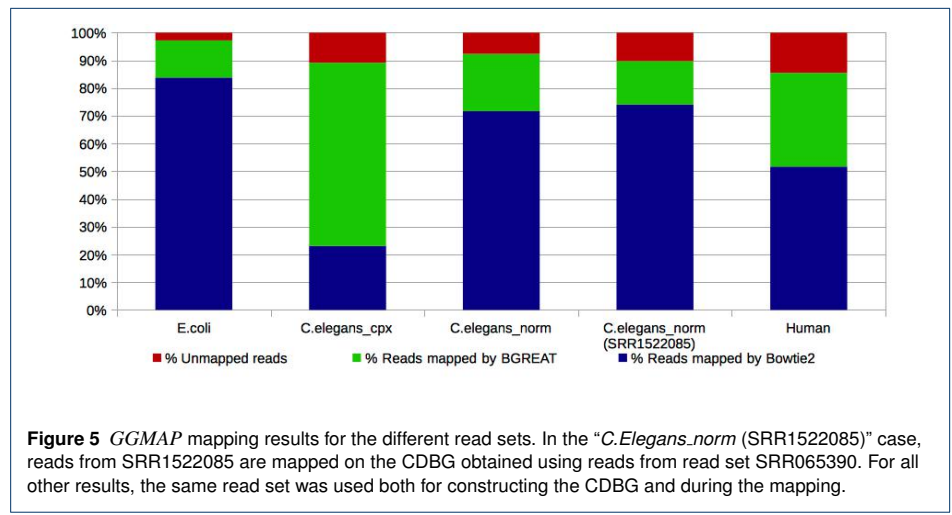

**Figure 5** *GGMAP* mapping results for the different read sets. In the "*C.Elegans_norm* (SRR1522085)" case, reads from SRR1522085 are mapped on the CDBG obtained using reads from read set SRR065390. For all other results, the same read set was used both for constructing the CDBG and during the mapping.

### 3.2 Graph mapping vs assembly mapping

We compared *GGMAP* to the popular approach consisting in mapping the reads to the reference contigs computed by an assembler. For testing this approach, for each of the three sets used, we first assembled them and then we mapped back the reads on the obtained set of contigs. We used two different assemblers, the widely used Velvet [22], and Minia [23], a memory efficient assembler based on Bloom filters. Finally, we used Bowtie2 for mapping the reads on the obtained contigs.

The results reported in Table 2 show that the number of reads mapped on assembled contigs is smaller than the one obtained with *GGMAP*. We obtained similar results in terms of number of reads mapped on the assemblies yielded by Velvet and Minia (see Table 2 of the Additional File). Let us emphasize that on the Human dataset, *GGMAP* maps 22 additional percents of reads on the graph than Bowtie2 does on the assembly.

We notice that the more complex the graph, the higher the advantage of mapping on the CDBG. This is due to the inherent difficulty of assembling with huge and highly branching graphs. This is particularly prominent in the results obtained on the artificially complex *C.elegans_cpx* CDBG.

We also highlight that our approach is resource efficient compared to most assembly processes. For instance, Velvet used more than 80 gigabytes of memory to compute the contigs for the *C. elegans* data set with $k = 31$. On this data set, our workflow used at most 4 GB memory (during $k$-mer counting). In terms of throughput, using *BGREAT* and then Bowtie2 on long unitigs is comparable to using Bowtie2 on contigs alone. See section 3.4 for more details about *GGMAP* performances.

### 3.3 Mapping on branching paths usefulness

Mapping the reads on branching paths of the graph is not equivalent to simply mapping the reads on unitigs. Indeed, at least 13% of reads (mapping reads SRR959239 on the *E.coli* DBG) and up to 66% of reads (mapping reads SRR065390 on *C.elegans_cpx* DBG) map on the branching paths of the graph (see Figure 5). These reads cannot be mapped when using only the set of unitigs as a reference. As expected, the more complex the graph, the larger the benefit of *BGREAT*'s approach. On the complex *C.elegans_cpx* graph, only 23% of reads can be fully mapped on unitigs, while 89% of them are mapped by additionally using *BGREAT*. On a simpler graph as *C.elegans_norm* the gap is smaller, but remains significant (72% vs 93%). Complete mapping results are shown in Table 3 of the Additional File.

*Non reflexive mapping on a CDBG*

The *GGMAP* approach is also suitable for mapping a distinct read set from the one used for constructing the DBG. We mapped another read set from *C.elegans* (SRR1522085) on the *C.elegans_norm* CDBG. Results in this situation are similar to those observed when performing reflexive mapping (i.e., when mapping the reads used to construct this graph): among 89% of mapped reads, 15% were mapped on branching paths of the graph (See Figure 5).

### 3.4 *GGMAP* performances

| CDBG Id | Mapped set (nb reads) | *BGREAT* Wall clock time | *BGREAT* CPU time | *BGREAT* Memory | BOWTIE2 Wall clock time | BOWTIE2 CPU time | BOWTIE2 Memory |
|---|---|---|---|---|---|---|---|
| *E.coli* | SRR959239 (5,128,790) | 28s | 1m40 | 19 MB | 1m17 | 3m53 | 29MB |
| *C.elegans_cpx* | SRR065390 (67,155,743) | 19m21 | 72m31 | 975MB | 8m12 | 33m | 1.66GB |
| *C.elegans_norm* | " | 13m03 | 51m28 | 336MB | 17m49 | 72m31 | 493MB |
| *C.elegans_norm* | SRR1522085 (22,509,110) | 1m54 | 7m13 | 336MB | 3m29 | 14m12 | 493MB |
| Human | SRR345593 SRR345594 (2,967,536,821) | 4h30 | 87h | 9.7GB | 4h38 | 90h15 | 21GB |

**Table 3** Time and memory footprints of *BGREAT* and BOWTIE2. Indicated wall clock times use four cores, except for the human samples for which 20 cores were used.

Table 3 presents *GGMAP* time and memory footprints. It shows that *BGREAT* is very efficient in terms of throughput while using moderate resources. Presented heuristics and implementation details allow *BGREAT* to scale up to real-world instances of the problem, being able to map millions of reads per CPU hour on a Human CDBG with a low memory footprint. *BGREAT* mapping is parallelized and can efficiently use dozens of cores.

### 3.5 *GGMAP* accuracy

To measure the impact of the read alignment heuristics, we forced the tool to explore exhaustively all potential alignment paths once a read is anchored on the graph. Results on the *E.coli* dataset show that the greedy approach is much faster than the exhaustive one (38× faster), while the mapping capacity is little impacted: the overall number of mapped reads increases by only 0.03% with the exhaustive approach. We thus claim that the choice of the greedy strategy is a satisfying trade-off.

| % Errors in simulated reads | Distance to optimum of *BGREAT* mapped reads (percentage) | | | | |
|---|---|---|---|---|---|
| | 0 | 1 | 2 | 3 | ≥ 4 |
| 0 | 100 | 0 | 0 | 0 | 0 |
| 0.1 | 99.31 | 0.52 | 0.09 | 0.04 | 0.04 |
| 0.2 | 98.79 | 0.91 | 0.21 | 0.07 | 0.02 |
| 0.5 | 97.2 | 2.17 | 0.41 | 0.17 | 0.05 |
| 1 | 94.88 | 3.72 | 0.92 | 0.41 | 0.07 |
| 2 | 90.85 | 6.43 | 1.79 | 0.83 | 0.1 |

**Table 4** *GGMAP* mapping results on simulated reads from the reference of the human chromosome 1 with default parameters. Results show the recall of *GGMAP* and the quality of *BGREAT* mapping, as represented by the "distance to optimum" value. For instance 94.88 % of the reads were mapped without error, 3.72% were mapped with a distance to the optimum of one etc. Due to approximate repeats in human chromosome 1, the reported distance to optimum is an upper bound.

To further evaluate the *GGMAP* accuracy, we assess the recall and mapping quality in the following experiment. We created a CDBG from Human chromosome 1 (hg19 version). Thus, each *k*-mer of the chromosome appears in the graph. Furthermore, from the same sequence, we simulated reads with distinct error rates (0%, 0.1%, 0.2%, 0.5%, 1% and 2%). For each error rate value, we generated one million reads. We evaluated the *GGMAP* results by mapping the simulated reads on the graph. As the graph is error free, except in some rare cases due to repetitions, the differences between a correctly mapped read and the path it maps to in the graph occur at erroneous positions of the read. If this is not the case, we say that the read is not mapped at its optimal position. Among the error free positions of a simulated read, the number of mismatches observed between this read and the mapped path is called the "*distance to optimal*". Results are reported in Table 4 together with the obtained recall (number of mapped reads over the number of simulated reads). Those results show the limits of *BGREAT* while mapping reads from divergent individuals. With 2% of substitutions in reads, only 90.85% of the reads are perfectly mapped. Nevertheless, with this divergence rate, 97.28% of reads are mapped at distance at most one from optimum. With over 99% of perfectly mapped reads, these results show that with the current sequencing characteristics, i.e. a 0.1% error rate, the mapping accuracy of *BGREAT* is suitable for most applications.

## 4 Discussion

We proposed a formal definition of the de Bruijn graph Read Mapping Problem (DB-GRMP) and proved its NP-completeness. We proposed a heuristic algorithm offering a practical solution. We developed a tool called *BGREAT* implementing this algorithm using a compacted de Bruijn graph (CDBG) as a reference.

From the theoretical viewpoint, the problem DBGRMP considers paths rather than walks in the graph. The current proof of its hardness does not seem to be adaptable to the cases of walks. A perspective is to extend the hardness result to that more general case.

We emphasize that our proposal does not enable genome annotation. It has been designed for applications aiming at a precise quantification of sequenced data, or a set of potential variations between the reads and the reference genome. In this context, it is essential to map as much reads as possible. Experiments show that a significant proportion of the reads (between ≈ 13% and ≈ 66% depending on the experiment) can be only mapped on branching paths of the graph. Hence, mapping only on the nodes of the graph or on assembled contigs is thus insufficient. This statement holds true when mapping the reads used for building the graph, but also with reads from a different experiment. Moreover, our results show that a potentially large number of reads (up to ≈ 32%) that are mapped on a CDBG cannot be mapped on a classical assembly.

With *GGMAP*, the mapping quality is very high: using Human chromosome 1 as a reference and reads with a realistic error rate (similar to that of Illumina technology), over 99% of the reads are correctly mapped. The same experiment also pointed out the limits of mapping reads on a divergent graph reference ($\geq 2\%$ substitutions): approximately 10% of the reads are mapped at a suboptimal position.

A weak point of *BGREAT* lies in its anchoring technique. Reads mapped with *BGREAT* must contain at least one exact $k-1$-mer that is an arc of the CDBG, *i.e.*, an overlap between two connected nodes. This may be a serious limitation when the original read set diverges greatly from the reads to be mapped. Improving the mapping technique may be done by using not only unitig overlaps as anchors at the cost of higher computational resources. Another solution may consist in using a smarter anchoring approach, like spaced seeds, which can accommodate errors in the anchor [24].

A natural extension consists in adapting *BGREAT* for mapping, on the CDBG obtained from short reads, the long (a few kilobases in average) and noisy reads produced by the third generation of sequencers, whose error rate reaches up to 15% (with mostly insertion and deletion errors for e.g. Pacific Biosciences technology). Such adaptation is not straightforward because of our seeding strategy, which requires long exact matches. The anchoring process must be very sensitive and very specific, while the mapping itself must implement a Blast-like heuristic or an alignment-free method. However, mapping such long reads on a DBG could be of interest for correcting these reads as in [13], or for solving repeats, if long reads are mapped on the walks (which main include cycles) of the DBG. Our NP-completeness proof only considers mapping on (acyclic) paths. Proving the hardness of the problem of mapping reads on walks of a DBG remains open.

Incidentally, using the same read set for constructing the CDBG and for mapping opens the way to major applications. Indeed, the graph and the exact location of each read on it may be used for **i/** read correction as in [15], by detecting differences between reads and the mapped area of the graph in which low support $k$-mers likely due to sequencing errors are absent, or for **ii/** read compression by recording additionally the mapping errors, or for **iii/** both correction and compression by conserving only for each read its mapping location on the graph.

Having for each read (used for constructing the graph or not) its location on the CDBG also provides the opportunity to design algorithms for enriching the graph, for instance enabling a quantification that is sensitive to local variations. This would be valuable for applications such as variant calling, analysis of RNA-seq variants [25], or of metagenomic reads [26].

Additionally, *BGREAT* results provide pieces of information for distant $k$-mers in the CDBG, about their co-occurrences in the mapped read data sets. This offers a way for the resolution, in the de Bruijn graph, of repeats larger than $k$. It could also allow to phase the polymorphisms and to reconstruct haplotypes.

## 5 Conclusion

A take home message is that read mapping can be significantly improved by mapping on the structure of an assembly graph rather than on a set of assembled contigs (respectively ≈22% and ≈ 32% of additional reads mapped for the Human and a complex *C.elegans* data sets). This is mainly due to the fact that assembly graphs retains more genomic information than assembled contigs, which also suffer from errors induced by the complexity of assembly. Moreover, mapping on a compacted De Bruijn Graph can be fast. The availability of *BGREAT* opens the door to its application to fundamental tasks such as read error correction, read compression, variant quantification, or haplotype reconstruction.

**Abbreviations**
DBG: De Bruijn Graph
CDBG: Compacted De Bruijn Graph
DBGRMP: De Bruijn Graph Read Mapping Problem
HPP: Hamiltonian Path Problem
FLATSP: Fixed Length Assymetric Travelling Salesman Problem
GRMP: Graph Read Mapping Problem

**Ethics approval and consent to participate**
Not applicable

**Availability of data and materials**
Our implementations are available at github.com/Malfoy/BGREAT In addition to the following pieces of information, Table 1 in Additional File presents the main characteristics of these datasets.
SRR959239 http://www.ncbi.nlm.nih.gov/sra/?term=SRR959239
SRR065390 http://www.ncbi.nlm.nih.gov/sra/?term=SRR065390
SRR1522085 http://www.ncbi.nlm.nih.gov/sra/?term=SRR1522085
SRR345593 and SRR345594 http://www.ncbi.nlm.nih.gov/sra/?term=SRR345593

**Competing interests**
The authors declare that they have no competing interests.

**Funding**
This work was funded by French ANR-12-BS02-0008 Colib'read project, by ANR-11-BINF-0002, and by a MASTODONS project.

**Author's contributions**
PP initiated the work and designed the study. AL, BC and ER designed the formalism and the proofs of NP-hardness. AL designed the algorithmic framework, implemented the *BGREAT* and performed the tests. All authors wrote and accepted the final version of the manuscript.

**Acknowledgements**
We would like to thank Yannick Zakowski, Claire Lemaitre and Camille Marchet for proofreading the manuscript and discussions.

**Additional files**
File name: $additionnal\_material.pdf$
Title: Read mapping on De Bruijn graphs additional file
Description: Three complementary tables are presented
- Main characteristics of data sets used in this study
- Assembly and mapping approach comparison
- Results of BGREAT on real read sets.

**Author details**
[1]IRISA Inria Rennes Bretagne Atlantique, GenScale team, Campus de Beaulieu, 35042 Rennes, France. [2]L.I.R.M.M., UMR 5506, Université de Montpellier et CNRS, 860 rue de St Priest, F-34392 Montpellier Cedex 5, France. [3]Institut Biologie Computationnelle, Université de Montpellier, F-34392 Montpellier, France.

**References**
1. Bradnam, K.R., Fass, J.N., *et al.*: Assemblathon 2: evaluating de novo methods of genome assembly in three vertebrate species. GigaScience **2**, 10 (2013). doi:10.1186/2047-217X-2-10. 1301.5406
2. Nagarajan, N., Pop, M.: Parametric Complexity of Sequence Assembly: Theory and Applications to Next Generation Sequencing. Journal of Computational Biology **16**(7), 897–908 (2009). doi:10.1089/cmb.2009.0005
3. Chaisson, M.J., Pevzner, P.A.: Short read fragment assembly of bacterial genomes. Genome Research **18**(2), 324–330 (2008). doi:10.1101/gr.7088808
4. Myers, E.W., Sutton, G.G., *et al.*: A whole-genome assembly of Drosophila. Science (New York, N.Y.) **287**(5461), 2196–2204 (2000). doi:10.1126/science.287.5461.2196
5. Myers, E.W.: The fragment assembly string graph. Bioinformatics **21**(Suppl 2), 79–85 (2005). doi:10.1093/bioinformatics/bti1114
6. Li, H., Durbin, R.: Fast and accurate short read alignment with Burrows-Wheeler transform. Bioinformatics **25**(14), 1754–1760 (2009). doi:10.1093/bioinformatics/btp324
7. Langmead, B., Salzberg, S.L.: Fast gapped-read alignment with Bowtie 2. Nature Methods **9**(4), 357–359 (2012). doi:10.1038/nmeth.1923. #14603
8. Lee, H., Schatz, M.C.: Genomic dark matter: the reliability of short read mapping illustrated by the genome mappability score. Bioinformatics **28**(16), 2097–2105 (2012). doi:10.1093/bioinformatics/bts330

9. Deshpande, V., Fung, E.D.K., Pham, S., Bafna, V.: Cerulean: A Hybrid Assembly Using High Throughput Short and Long Reads. In: Lecture Notes in Computer Science vol. 8126 LNBI, pp. 349–363 (2013). doi:10.1007/978-3-642-40453-5_27
10. Ribeiro, F.J., Przybylski, D., Yin, S., Sharpe, T., Gnerre, S., Abouelleil, A., Berlin, A.M., Montmayeur, A., Shea, T.P., Walker, B.J., Young, S.K., Russ, C., Nusbaum, C., MacCallum, I., Jaffe, D.B.: Finished bacterial genomes from shotgun sequence data. Genome Research **22**(11), 2270–2277 (2012). doi:10.1101/gr.141515.112
11. Pevzner, P.A., Tang, H., Waterman, M.S.: An Eulerian path approach to DNA fragment assembly. Proceedings of the National Academy of Sciences **98**(17), 9748–9753 (2001). doi:10.1073/pnas.171285098
12. Bankevich, A., Nurk, S., Antipov, D., Gurevich, A.A., Dvorkin, M., Kulikov, A.S., Lesin, V.M., Nikolenko, S.I., Pham, S., Prjibelski, A.D., Pyshkin, A.V., Sirotkin, A.V., Vyahhi, N., Tesler, G., Alekseyev, M.A., Pevzner, P.A.: SPAdes: A New Genome Assembly Algorithm and Its Applications to Single-Cell Sequencing. Journal of Computational Biology **19**(5), 455–477 (2012). doi:10.1089/cmb.2012.0021
13. Salmela, L., Rivals, E.: LoRDEC: accurate and efficient long read error correction. Bioinformatics **30**(24), 3506–3514 (2014). doi:10.1093/bioinformatics/btu538
14. Yang, X., Chockalingam, S.P., Aluru, S.: A survey of error-correction methods for next-generation sequencing. Briefings in bioinformatics **14**(1), 56–66 (2013). doi:10.1093/bib/bbs015
15. Benoit, G., Lavenier, D., Lemaitre, C., Rizk, G.: Bloocoo, a memory efficient read corrector. In: European Conference on Computational Biology (ECCB) (2014)
16. Wang, M., Ye, Y., Tang, H.: A de Bruijn Graph Approach to the Quantification of Closely-Related Genomes in a Microbial Community. Journal of Computational Biology **19**(6), 814–825 (2012). doi:10.1089/cmb.2012.0058
17. Huang, L., Popic, V., Batzoglou, S.: Short read alignment with populations of genomes. Bioinformatics **29**(13), 361–370 (2013). doi:10.1093/bioinformatics/btt215
18. Dilthey, A., Cox, C., Iqbal, Z., Nelson, M.R., McVean, G.: Improved genome inference in the MHC using a population reference graph. Nature Genetics **47**(6), 682–688 (2015). doi:10.1038/ng.3257
19. Holley, G., Peterlongo, P.: Blastgraph: Intensive approximate pattern matching in sequence graphs and de-bruijn graphs. In: Stringology, pp. 53–63 (2012)
20. Karp, R.M.: Reducibility Among Combinatorial Problems. In: 50 Years of Integer Programming 1958-2008, pp. 219–241. Springer, Berlin, Heidelberg (2010). doi:10.1007/978-3-540-68279-0_8. http://link.springer.com/10.1007/978-3-540-68279-0_8
21. Chikhi, R., Limasset, A., *et al.*: On the representation of de bruijn graphs. In: Lecture Notes in Computer Science, vol. 8394 LNBI, pp. 35–55 (2014). doi:10.1007/978-3-319-05269-4-4
22. Zerbino, D.R., Birney, E.: Velvet: Algorithms for de novo short read assembly using de Bruijn graphs. Genome Research **18**(5), 821–829 (2008). doi:10.1101/gr.074492.107
23. Chikhi, R., Rizk, G.: Space-efficient and exact de Bruijn graph representation based on a Bloom filter. Algorithms for Molecular Biology **8**(1), 22 (2013). doi:10.1186/1748-7188-8-22
24. Vroland, C., Salson, M., Touzet, H.: Lossless Seeds for Searching Short Patterns with High Error Rates. In: Combinatorial Algorithms - 25th International Workshop, IWOCA 2014, Duluth, MN, USA, October 15-17, 2014, pp. 364–375 (2014)
25. Sacomoto, G.A., Kielbassa, J., Chikhi, R., Uricaru, R., Antoniou, P., Sagot, M.-F., Peterlongo, P., Lacroix, V.: Kissplice: de-novo calling alternative splicing events from rna-seq data. BMC bioinformatics **13**(Suppl 6), 5 (2012)
26. Ye, Y., Tang, H.: Utilizing de bruijn graph of metagenome assembly for metatranscriptome analysis. arXiv preprint arXiv:1504.01304 (2015)